\documentclass[twocolumn,trackchanges]{aastex631}

\usepackage{CJK}
\newcommand{\petit}{\texttt{petitRADTRANS}}

\newcommand{\teff}{T$_{\rm eff}$}

\newcommand{\kms}{$\rm km\,s^{-1}$}
\newcommand{\gj}{GJ~436}
\newcommand{\gjb}{GJ~436~b}
\newcommand{\kp}{$K_{\rm p}$}
\newcommand{\dvsys}{$\Delta v_{\rm sys}$}
\newcommand{\kpvsys}{$K_{\rm p}-\Delta v_{\rm sys}$}
\newcommand{\caltech}{Department of Astronomy, California Institute of Technology, Pasadena, CA 91125, USA}
\newcommand{\gps}{Division of Geological \& Planetary Sciences, California Institute of Technology, Pasadena, CA 91125, USA}
\newcommand{\ucsc}{Department of Astronomy \& Astrophysics, University of California, Santa Cruz, CA95064, USA}
\newcommand{\keck}{W. M. Keck Observatory, 65-1120 Mamalahoa Hwy, Kamuela, HI 96743, USA}
\newcommand{\ucla}{Department of Physics \& Astronomy, 430 Portola Plaza, University of California, Los Angeles, CA 90095, USA}
\newcommand{\jpl}{Jet Propulsion Laboratory, California Institute of Technology, 4800 Oak Grove Dr.,Pasadena, CA 91109, USA}
\newcommand{\ucsd}{Department of Astronomy \& Astrophysics, University of California San Diego, La Jolla, CA 92093, USA}

\newcommand{\northwestern}{Center for Interdisciplinary Exploration and Research in Astrophysics (CIERA) and Department of Physics and Astronomy,
Northwestern University, Evanston, IL 60208, USA}
\newcommand{\arizona}{James C. Wyant College of Optical Sciences, University of Arizona,
Meinel Building 1630 E. University Blvd., Tucson, AZ 85721, USA}

\received{\today}
\shorttitle{Possible HRCCS detection of GJ 436 b}
\shortauthors{Finnerty et al.}
\graphicspath{{./}{}}
\begin{document}
\begin{CJK*}{UTF8}{gbsn}

\title{Possible stratospheric emission in the warm Neptune GJ 436 b from high-resolution spectroscopy}

\correspondingauthor{Luke Finnerty}
\email{lfinnert@umich.edu}

\author[0000-0002-1392-0768]{Luke Finnerty}
\affiliation{\ucla}

% Fitzgerald
\author[0000-0002-0176-8973]{Michael P. Fitzgerald}
\affiliation{\ucla}

% Xuan
\author[0000-0002-6618-1137]{Jerry W. Xuan}
\affiliation{\caltech}

\author[0000-0002-1583-2040]{Daniel Echeverri}
\affiliation{\caltech}

% Jovanovic
\author[0000-0001-5213-6207]{Nemanja Jovanovic}
\affiliation{\caltech}

% Mawet
\author{Dimitri Mawet}
\affiliation{\caltech}
\affiliation{\jpl}

\author{Geoffrey A. Blake}
\affiliation{\gps}

\author[0000-0002-6525-7013]{Ashley Baker}
\affiliation{\caltech}

\author{Randall Bartos}
\affiliation{\jpl}

% % Calvin
\author[0000-0003-4737-5486]{Benjamin Calvin}
\affiliation{\caltech}
\affiliation{\ucla}

% Cetre
\author{Sylvain Cetre}
\affiliation{\keck}

% Delorme
\author[0000-0001-8953-1008]{Jacques-Robert Delorme}
\affiliation{\keck}
\affiliation{\caltech}

% Doppman
\author{Greg Doppmann}
\affiliation{\keck}

\author[0000-0001-9708-8667]{Katelyn Horstman}
\affiliation{\caltech}
\altaffiliation{NSF Graduate Research Fellow}

\author[0000-0002-5370-7494]{Chih-Chun Hsu}
\affiliation{\northwestern}

% Inglis
\author{Julie Inglis}
\affiliation{\caltech}

% Liberman
\author[0000-0002-4934-3042]{Joshua Liberman}
\affiliation{\caltech}
\affiliation{\arizona}

% Lopez
\author[0000-0002-2019-4995]{Ronald A. L\'opez}
\affiliation{\ucla}

% Morris
\author{Evan Morris}
\affiliation{\ucsc}

% Pezzato
\author{Jacklyn Pezzato-Rovner}
\affiliation{\caltech}

% Ruffio
\author[0000-0003-2233-4821]{Jean-Baptiste Ruffio}
\affiliation{\ucsd}

% Sappey
\author[0000-0003-1399-3593]{Ben Sappey}
\affiliation{\ucsd}

% Schofield
\author{Tobias Schofield}
\affiliation{\caltech}

% Skemer
\author{Andrew Skemer}
\affiliation{\ucsc}

% Wallace
\author[0000-0001-5299-6899]{J. Kent Wallace}
\affiliation{\jpl}

% Wallack
\author[0000-0003-0354-0187]{Nicole L. Wallack}
\affiliation{Earth and Planets Laboratory, Carnegie Institution for Science, Washington, DC 20015, USA}

% Wang
\author[0000-0003-0774-6502]{Jason J. Wang (王劲飞)}
\affiliation{\northwestern}

% Wang (王吉)
\author[0000-0002-4361-8885]{Ji Wang (王吉)}
\affiliation{Department of Astronomy, The Ohio State University, 100 W 18th Ave, Columbus, OH 43210 USA}

% Xin
\author[0000-0002-6171-9081]{Yinzi Xin}
\affiliation{\caltech}

\begin{abstract}
We present high spectral resolution $L$ band (2.91--3.85 $\mu$m) observations of the warm Neptune \gjb\ from Keck II/KPIC. KPIC's single-mode fiber feed reduces the $L$ band sky background by a factor of 100, significantly improving sensitivity compared to a seeing-limited spectrometer and enabling a tentative ($\rm SNR = 3-4$) cross-correlation detection of \gjb\ with a thermally inverted atmospheric model. In contrast with recent results from \textit{JWST} and high-resolution transmission spectroscopy, our retrieval analysis prefers the presence of H$_2$O, and possibly CH$_4$, molecular features in emission. The broad-band continuum flux associated with the maximum-likelihood model is substantially higher than expected based on both the $\sim670\rm\ K$ equilibrium temperature of \gjb\ and previous results from low-resolution spectroscopy. We demonstrate that the loss of continuum information during the processing of high-resolution spectra makes our analysis effectively insensitive to the absolute continuum level of the planet, and that scaling the maximum-likelihood model to match the broad-band flux measured from low-resolution observations of \gjb\ results in a detection of similar strength in cross-correlation. These results could be explained by a thermal inversion arising above a haze layer in the upper atmosphere of \gjb. Further observations, ideally post-eclipse in order to break the \kpvsys\ degeneracy, are needed to clarify this possible detection. This work demonstrates the potential of $L$ band high-resolution spectroscopy for characterizing significantly smaller and cooler exoplanets compared with hot Jupiters. 

\end{abstract}



\section{Introduction} \label{sec:intro}
\gjb\ is the archetypal example of the warm Neptune exoplanet population. At the time of its discovery in 2004, it was the lowest-mass exoplanet known, and only the second exoplanet detected around an M-dwarf \citep{butler2004}. \citet{gillon2007} subsequently reported the first transit detection of \gjb. As a warm Neptune around a bright host star, \gjb\ quickly became a popular target for infrared characterization studies. Secondary eclipse spectroscopy from \textit{Spitzer} at 8 $\mu$m was quickly used by \citet{deming2007} to confirm the non-zero orbital eccentricity of \gjb, with similar results reported by \citet{demory2007}. 

Transmission spectroscopy observations from both space and the ground have consistently yielded featureless spectra for \gjb, suggestive of either high atmospheric metallicity or significant clouds/hazes around 1-10 mbar muting the transmission spectrum \citep{pont2009, gibson2011, knutson2014, lothringer2018, grasser2024}. Stellar activity in the M-dwarf primary further complicates the interpretation of transmission spectra, but was helpful in suggesting a misaligned, nearly polar orbit \citep{knutson2011}, which has been subsequently confirmed \citep{bourrier2022}. UV transmission observations have had more success in making clear detections, detecting an extended hydrogen envelope escaping from \gjb\ \citep{ehrenreich2015}.

Emission observations are less impacted by clouds and hazes than transmission observations, but a consistent interpretation of \gjb\ observations has remained elusive. \citet{stevenson2010} obtained broad-band secondary eclipse measurements in six \textit{Spitzer} bandpasses, finding a very high planet flux at 3.6 $\mu$m and low flux at 4.5$\mu$m, suggesting a very high CO/CH$_4$ ratio as a result of disequilibrium chemistry. A reanalysis of \textit{Spitzer} data by \citet{lanotte2014} found somewhat shallower eclipse depths at these wavelengths, still consistent with a high CO/CH$_4$ ratio but compatible with somewhat lower metallicities ($\sim50\times$ solar compared with $\sim200\times$ solar). Subsequent \textit{Spitzer} observations confirmed the lower 3.6 $\mu$m and 4.5 $\mu$m eclipse depths, and retrievals on the \textit{Spitzer} photometry preferred a non-inverted atmosphere with a metallicity several hundred times greater than solar \citep{morley2017}.

Recent low-resolution spectroscopy observations from \textit{JWST} have found a much lower $3.6\ \mu \rm m$ flux than expected based on the \textit{Spitzer} photometry \citep{mukherjee2025}. Re-analysis of the \textit{Spitzer} photometry by \citet{mukherjee2025} suggests the anomalously high $3.6\ \mu \rm m$ point may be a detector artifact. Depending on model assumptions, the \textit{JWST} spectra could be explained by a cloudy, very high-metallicity ($>300\times$ solar) atmosphere, or a clear atmosphere with relatively low ($>80\times$ solar) metallicity and low internal temperature \citep{mukherjee2025}.

In this paper, we present observations of \gjb\ in $L$ band emission with Keck II/KPIC high-resolution spectroscopy covering $2.91-3.85\rm\ \mu m$, with significant gaps. By observing in thermal emission at high spectral resolution, our observations are sensitive to weak spectral features arising above a cloud/haze layer, which may be missed in transmission or at lower spectral resolution. Section \ref{sec:obs} describes the details of our observations, reduction procedure, and retrieval analysis. Results of this analysis are presented in Section \ref{sec:res}, and discussed in the context of previous results in Section \ref{sec:disc}. Section \ref{sec:conc} concludes.

 \begin{deluxetable}{ccc}
    \tablecaption{Stellar and planetary properties for the \gj\ system.} 
    \label{tab:props}
    \tablehead{\colhead{Property} & \colhead{Value} & \colhead{Ref.}}
    \startdata
        & \textbf{\gj} & \\
        \hline
        RA & 11:42:12.13 &  \citet{gaiaedr3} \\
        Dec & +26:42:11 &  \citet{gaiaedr3} \\
        Sp. Type & M3V & \citet{kirkpatrick} \\
        $K_{\rm mag}$ & $6.07\pm0.02$ & \citet{cutri2003} \\
        Mass & $0.44\pm0.01\rm\ M_\odot$ & \citet{rosenthal2021}  \\
        Radius & $0.42\pm0.01\rm\ R_\odot$ & \citet{rosenthal2021} \\
        \teff & $3600\pm40$ K & \citet{rosenthal2021} \\
        $\log g$ [cgs] & $4.84\pm0.01$ & \citet{rosenthal2021}  \\
        $v\sin i$ & $0.33\pm0.1$ \kms & \citet{bourrier2022} \\
        $v_{\rm rad}$ & $9.6\pm0.3$ \kms & \citet{nidever2002}  \\
        $\rm [Fe/H]$ & $0.1\pm0.08$ & \citet{rosenthal2021} \\
        % $\rm [C/H]$ &\\
        % $\rm [O/H]$ & \\
        % C/O &  & \\
        \smallskip \\
        \hline
         & \textbf{\gjb} & \\
        \hline
        Period & $2.64389753\pm1\times10^{-7}$  days & \citet{kokori23} \\
        $\rm T_{\rm trans}$ & $\rm JD\ 2454873.01582\pm4\times10^{-5} $ & \citet{mukherjee2025}   \\
        $a$ & $0.0285\pm0.0002$ AU & \citet{rosenthal2021}\\
        $e$ & $0.163\pm0.003$ & \citet{mukherjee2025} \\
        $\omega$ & $327.3\pm1.5^\circ$  & \citet{mukherjee2025} \\
        $i$ & $86.70^\circ\pm0.03^\circ$ & \citet{kokori23} \\
        $K_{\rm p}$ & 117 \kms  & Est. \\
        Mass & $0.067\pm0.002\rm\ M_J$ & \citet{rosenthal2021}\\
        Radius & $0.37\pm0.01 \rm\ R_J$ & \citet{turner2016} \\
        $\rm T_{\rm eq}$ &  670 K & Est. \\
        % C/O &  & This work \\
        % $\rm [C/H]$ &  & This work \\
        % $\rm [O/H]$ & & This work \\
        % $v\sin i$ & & This work 
    \enddata
\end{deluxetable}

\section{Observations and Data Reduction}\label{sec:obs}

\subsection{Observations}
\gjb\ was observed using Keck II/KPIC \citep{nirspec, nirspecupgrade, nirspecupgrade2, kpic, echeverri2022, kpicII} on UT 2024 May 22 from 5:43 to 9:50. The observations were taken close to secondary eclipse, at an orbital phase of 0.42--0.48, during which time the nominal projected orbital velocity based on the orbital parameters in Table \ref{tab:props} changed from 89 \kms\ to 63 \kms, accounting for the non-zero orbital eccentricity. Due to the high and variable slit background in the $L$ band, observations were taken in an ABBA nodding pattern using science fibers 2 and 4, rather than the staring mode used for $K$-band observations \citep[e.g.][]{finnerty2025a}. We used an exposure time of 60 seconds in order to keep the entire detector in the linear regime despite significant thermal background at the red end of the observed spectrum, obtaining 102 frames on science fiber 2 and 104 frames on science fiber 4

\gj\ was observed starting just before transiting zenith, with the airmass changing from 1.01 to 1.63 over the observations.  Conditions were good and generally stable over the entire observation sequence, aside from a $\sim30$ minute patch of poor AO correction around 8:00. Outside of that period, typical counts gradually decreased by $\sim20\%$ over the course of the observations, roughly tracking the airmass. The entire time series for both science fibers was used for the analysis.

\subsection{Data Reduction}

The data were reduced using the modified version of the KPIC pipeline\footnote{\href{https://github.com/kpicteam/kpic_pipeline/}{https://github.com/kpicteam/kpic\_pipeline/}} described in \citet{finnerty2025a, finnerty2025b}, which uses a variable Gaussian-Hermite model of the trace profile for both flux extraction and the instrumental line-spread function (LSF). While \citet{finnerty2025a, finnerty2025b} used afternoon calibration frames for background subtraction, we found that the $L$ band slit background is significantly different between daytime calibrations and science operations, and that the background level varies slowly over the course of the night as a result of changes in temperature. We therefore took our observations in an ABBA pattern and performed an A--B subtraction to remove the thermal background. 

For wavelength calibration, we observed HIP 62944 before observations of \gj\ and observed HIP 81497 after. The limited number of existing KPIC $L$ band observations necessitated a more manual approach to wavelength calibration compared with previous $K$ band observations. We used the NIRSPEC Echelle Format Simulator (EFS) to obtain a rough estimate of the wavelength solution for each order. This was then refined by manually adjusting coefficients of a second-order polynomial to match a stellar $\times$ telluric model to the data. The resulting initial wavelength solution was then used as a starting point for the KPIC DRP wavelength calibration script. Finally, this wavelength solution was checked against the stellar $\times$ telluric model and refined by manually tuning coefficients to a sum of the first four Legendre polynomials ($L_{0-3}$). 

Optimal extraction weights and the line spread function (LSF) model were fit using the 5th-order Gaussian-Hermite model previously described in \citet{finnerty2025b, finnerty2025c}. We used the sum of the full stack of science frames for LSF fitting in order to maximize the total signal-to-noise for the fitting. Following \citet{finnerty2025a}, we apply a stretching factor of 1.14 to the LSF compared with the extraction weights to account for the asymmetry described in \citet{Finnerty2022}. The Full Width at Half Maximum (FWHM) of the resulting LSF corresponds to a spectral resolution $R = \lambda/\Delta\lambda = 29,000$ in the 2.91  $\mu$m order, decreasing to $21,100$ in the 3.8 $\mu$m order. These correspond to a velocity resolution of $10-14$ \kms, respectively. The nominal planet velocity shift of 26 \kms\ is therefore close to the minimum required for HRCCS analysis.  

Of the eight spectral orders falling on the detector, seven were usable, spanning 2.91--3.85 $\mu$m, with significant gaps between orders. The eighth, bluest order was strongly impacted by telluric absorption and is not suitable for further analysis. Subsequent KPIC $L$ band observations \citep[e.g.][]{finnerty2025c} have used a different grating/cross-disperser configuration spanning 3.03--4.06$\mu$m. The extracted SNR for each 60 second exposure in regions of low telluric absorption is strongly chromatic, ranging from $\sim30$ in the bluest order to $\sim100$ in the reddest order. 

\subsection{Atmospheric Retrieval}

We use the retrieval pipeline described in \citet{finnerty2023, finnerty2024, finnerty2025a, finnerty2025b}, using \petit\ for radiative transfer \citep{prt:2019, prt:2020, Nasedkin2024} and the \citet{brogi2019} log-likelihood mapping. We added an additional masking step in the pre-retrieval data processing in order to better handle the saturated telluric features in $L$. After median-dividing each spectrum to establish a consistent continuum level, any wavelength channel with $<70\%$ telluric transmission is masked. We then mask the $3\%$ of the remaining wavelength channels with the highest variance, before computing the time series median and dividing the entire time series for each order by the median spectrum. Principal Component Analysis (PCA) along the time axis is then used to remove the temporally varying tellurics and instrument drifts, which are not removed by the median division. We performed the retrieval analysis omitting 2, 4, and 6, and 8 principal components (PCs) from the time series. We assess the impact of varying the number of omitted PCs below. The omitted principal components are saved and added to the forward model in order to replicate the impact of the PCA on the underlying planet signal, as described in \citet{line2021} and previously used for KPIC HRCCS analysis \citep{finnerty2024, finnerty2025a, finnerty2025b}. 

For the pressure-temperature ($P-T$) profile, we use the \citet{guillot2010} parameterization following \citet{finnerty2025a, finnerty2025b}, with four free parameters corresponding to the log infrared opacity ($\log \kappa$), log of the infrared/optical opacity ratio ($\log \gamma$), intrinsic temperature ($\rm T_{int}$), and equilibrium temperature ($\rm T_{eq}$). We do not include $\log g$ as a free parameter, as the $P-T$ profile implementation we use results in a direct correlation between $\log g$ and $\log \kappa$ \citep[see][for details]{finnerty2026a}. We instead set $\log g$ to the value for the nominal planet mass and radius listed in Table \ref{tab:props}. We include a gray cloud deck, as multiple studies have found evidence for significant clouds impacting the spectrum of \gjb. We fit for the cloud deck pressure, cloud opacity, and settling parameter $f_{\rm SED}$.  

As the orbital eccentricity of \gjb\ is significantly non-zero, we compute the projected orbital velocity of the planet using the full orbital solution reported by \citet{mukherjee2025}. For each frame, the planet velocity is given by:

\begin{equation}
    v_{pl}(t) = K_p(\cos(f(t)+\omega) + e\cos(\omega)) + v_{rad} - v_{bary}(t) + \Delta v_{sys}
\end{equation}

Where \kp\ and \dvsys\ are free parameters in the retrieval, and $f(t)$ is the true anomaly at frame time $t$, which is computed from the mean anomaly by solving the Kepler equation. The values of $\omega$, $e$, and $v_{rad}$ are fixed to those listed in Table \ref{tab:props}, and $v_{bary}(t)$ is the barycentric velocity at each frame. 

We fit for vertically constant abundances of H$_2$O, CH$_4$, NH$_3$, H$_2$S, HCN, SO$_2$ and H$_2$, using the default opacity tables included with \petit. While previous analyses using this pipeline have fit abundances as mass-mixing ratios \citep{finnerty2023, finnerty2024, finnerty2025a, finnerty2025b}, we have switched to fitting volume-mixing ratios to be more consistent with the broader literature. For H$_2$O, we used the opacities based on the \citet{polyansky2018} POKAZATEL linelist. For CH$_4$, we use the \citet{hargreaves2020} linelist. For NH$_3$, we use the \citet{yurchenko2011} linelist. H$_2$S opacity is based on the \citet{azzam2016} linelist, HCN uses \citet{harris2006}, and SO$_2$ uses \citet{underwood2016}, imported from DACE \citep{grimm2015, grimm2021}. H$_2$ opacity is from \citet{rothman2013}. We also include $\rm H_2 - H_2$ CIA opacity from \citet{borysow2001, borysow2002} and $\rm H_2-He$ opacity from \citet{borysow1988, borysow1989a, borysow1989b}. 

We use a stellar model from the PHOENIX libarary \citep{phoenix} with $\rm T_{eff} = 3600$ K, $\log g = 4.5$, and $[\rm Fe/H] = 0$. We rotationally broaden this model to $v\sin i = 0.33$\ \kms\ using the algorithm described by \citet{carvalho2023}. We do not include the change in stellar velocity over the observations in our forward model. This change is only a few hundred meters per second, much smaller than the intrument resolution, and arises from a combination of the changing barycentric velocity and the radial velocity pull of the planet. Neglecting the stellar velocity change is equivalent to assuming a smooth stellar model, similar to \citet{line2021}. In the case of \gj, which has significant spectral features, this may result in model mismatch near the stellar reference frame, but limits the impact of inaccuracies in the stellar model otherwise. The nominal planetary radial velocity track is $>4\times$ the spectral resolution from the host star reference frame throughout the observation, and we therefore expect correlation with residuals in the stellar reference frame to have minimal impact on our analysis.

The \citet{brogi2019} log-likelihood also includes a multiplicative scaling factor applied to the forward model. This factor can account for errors in the star/planet radius radio or errors in the stellar model which bias the resulting $F_p/F_s$ in the forward model. However, over a narrow bandpass a multiplicative scaling factor can become denenerate with the $P-T$ profile, which also effectively scales the strength of planet lines. We mitigate this by using a log-normal prior with a mean of 1, similar to \citet{finnerty2025b}. 

The sampling is performed using the \texttt{PyMultiNest} \citep{buchner2014} wrapper for \texttt{MultiNest} \citep{feroz2008, feroz2009, feroz2019}. We used 1200 live points and a convergence criteria of $\Delta \log z < 0.01$, consistent with our previous retrievals \citep{finnerty2023, finnerty2024, finnerty2025a, finnerty2025b, finnerty2025c}. 

\section{Results}\label{sec:res}

\subsection{Retrieval analysis}
\begin{deluxetable*}{ccccccc}
    \tablecaption{List of parameters, priors, and results for atmospheric retrievals.}
    \tablehead{\colhead{Name} & Symbol  & \colhead{Prior} & \colhead{Retrieved Max-L} & \colhead{Retrieved Median} & \colhead{Inverted test} & \colhead{Non-inverted test}} 
    \startdata
        log infrared opacity [$\rm cm^{2} g^{-1}$] & $\log \kappa$ &  Uniform($-4, 2$)    &  $-1.3$ & $-1.2^{+1.3}_{-1.3}$ & $-0.5$ & $-0.5$ \\
        log infrared/optical opacity & $\log \gamma$  &               Uniform($-4, 4$)    &  $3.5$ & $2.3^{+0.9}_{-0.9}$ & $1.2$ & $-1.2$ \\
        Intrinsic temperature [K] & $\rm T_{int}$ &                   Uniform($10,500$)   & $350$ & $160^{+160}_{-100}*$ & $150$ & $150$ \\
        Equilibrium temperature [K] & $\rm T_{equ}$ &                 Uniform($100,2400$) & $570$ & $1010^{+540}_{-410}$ & $600$ & $600$ \\
        Cloud settling parameter & $\log f_{\rm SED}$ &               Uniform($-2,2$)     & $-0.2$ & $0.0^{+1.2}_{-1.3}*$ & $-1$ & $-1$ \\ 
        Cloud opacity & $\log \kappa_{\rm cloud}$ &                   Uniform($-8,3$)     & $-6.1$ & $-2.4^{+3.4}_{-3.4}*$ & $0$ & $0$ \\ 
        Cloud deck pressure & $\rm \log P_{cloud}$ &                  Uniform($-6,-1$)    & $-4.5$ & $-3.5^{+1.5}_{-1.5}*$ & $-3$ & $-3$ \\ 
        $K_{\rm p}$ offset [\kms] & $\Delta K_{\rm p}$  &             Uniform($-60, 60$)  & $58.0$ & $50.0^{+6.0}_{-13.0} (> 13)$ & $0$ & $0$ \\
        $v_{\rm sys}$ offset [\kms] & \dvsys\ &                       Uniform($-40,40$)   & $-37.6$ & $-32.5^{+8.6}_{-4.4} (<-8 )$ & $0$ & $0$ \\
        Rotational velocity [\kms] & $v_{\rm rot}$ &                  Uniform($0,15$)     & $0.3$ & $3.0^{+2.7}_{-1.9} (< 8)$  & $3$ & $3$ \\
        log H$_2$O volume-mixing ratio & log H$_2$O  &                Uniform($-12, -0.3$) & $-2.4$ & $-3.3^{+1.3}_{-1.4}$ & $-1.2$ & $-1.6$ \\
        log CH$_4$ volume-mixing ratio & log CH$_4$ &                 Uniform($-12, -0.3$) & $-3.2$ & $-7.0^{+2.8}_{-3.1} (< -3.0)$ & $-1.3$ & $-1.7$ \\
        log NH$_3$ volume-mixing ratio & log NH$_3$ &                 Uniform($-12, -1$)   & $-6.2$ & $-7.7^{+2.8}_{-2.7} (< -3.5)$ &  $-4.0$ & $-5.4$ \\
        log H$_2$S volume-mixing ratio & log H$_2$S &                 Uniform($-12, -1$)   & $-2.4$ & $-7.1^{+3.1}_{-3.1}* $ & $-2.7$ & $-2.7$ \\
        log HCN volume-mixing ratio & log HCN &                       Uniform($-12, -1$)   & $-4.2$ & $-8.0^{+2.6}_{-2.4} (< -4.0)$ & $-12$ & $-9$ \\
        log SO$_2$ volume-mixing ratio & log SO$_2$ &                 Uniform($-12, -1$)   & $-3.5$ & $-6.6^{+3.3}_{-3.4}*$ & $-3.2$ & $-3.2$ \\
        log H$_2$ volume-mixing ratio & $\log \rm H_2$ &              Uniform($-0.4,-0.05$) & $-0.1$ & $-0.2^{+0.1}_{-0.1}*$ & $-0.3$ & $-0.3$ \\
        Scale factor & scale &                                        LogNormal($0,0.1$) & $0.07$ & $0.04^{+0.09}_{-0.09}$  & $0$ & $0$
    \enddata 
    \tablecomments{Parameters are listed for the four component retrieval. The full corner plots are included in Appendix \ref{app:corner}. The error bars on the retrieved medians correspond to the 68$\% / 1\sigma$ confidence interval. Parameters which were not constrained are marked with a $*$, and limits are given at 95\% confidence. In addition to these priors, we required that the atmospheric temperature stay below 3500 K at all pressure levels.}
    \label{tab:priors}
\end{deluxetable*}

Priors, maximum-likelihood parameters, and retrieved median parameters with $\pm34$\% confidence intervals are presented in Table \ref{tab:priors} for the four component retrieval. Full corner plots for all four retrievals are included in Appendix \ref{app:corner}.  The cross-correlation time series for each fiber with the maximum-likelihood model is shown in Figure \ref{fig:vtracks}, and the \kpvsys\ for the maximum-likelihood model for each fiber and varying number of omitted principal components is shown in Figure \ref{fig:kpvsys_pcs}. The maximum-likelihood model produces a weak cross-correlation peak in both science fibers in Figure \ref{fig:vtracks} at $\rm SNR\sim2-3$. Figure \ref{fig:kpvsys_pcs} shows similar results for the \kpvsys\ diagrams, with $\rm SNR\sim3-4$ after combining the two science fibers. These results are consistent with a marginal detection of \gjb. 

The four-component retrieval returns a constrained posterior. The retrieved velocity parameters run against the upper end of the prior, but are consistent with the offset in the cross-correlation space seen in Figure \ref{fig:kpvsys_pcs}. The retrieved $P-T$ profile has a strong thermal inversion, cloud parameters are unconstrained, and weak upper bounds are returned for all species except H$_2$O The posterior from the six and eight component retrievals are generally consistent with the four component retrieval. In contrast, the retrieved posterior omitting two principal components is bi-modal, with one mode corresponding to the posterior mode obtained in the retrievals omitting more PCs (though with a stronger preference for CH$_4$ emission), and a second, weaker mode dominated by H$_2$S absorption. The evolution of the posterior as more components are omitted and the implausibility of a H$_2$S-dominated atmosphere suggest this secondary mode is a result of uncorrected residuals in the spectroscopic time series, and motivates our adoption of the four-component case as fiducial. Previous analyses of KPIC observations have preferred omitting 4-6 components, with minimal impacts on the retrieved posterior for up to 8 components \citep[e.g.][]{finnerty2023, finnerty2024, finnerty2025b}, which is consistent with the results we obtain for \gjb\ in these data. The absolute flux level of the retrieved planet model is significantly higher than expected based on recent \textit{JWST} results, which we discuss in Section \ref{sec:disc}. 

\subsection{Cross-correlation analysis}

\begin{figure*}
    \centering
    \includegraphics[width=0.45\linewidth]{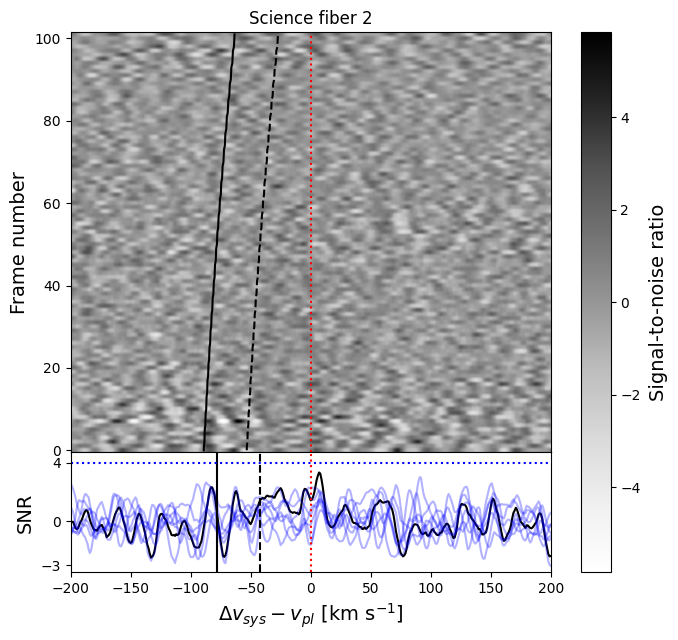}
    \includegraphics[width=0.45\linewidth]{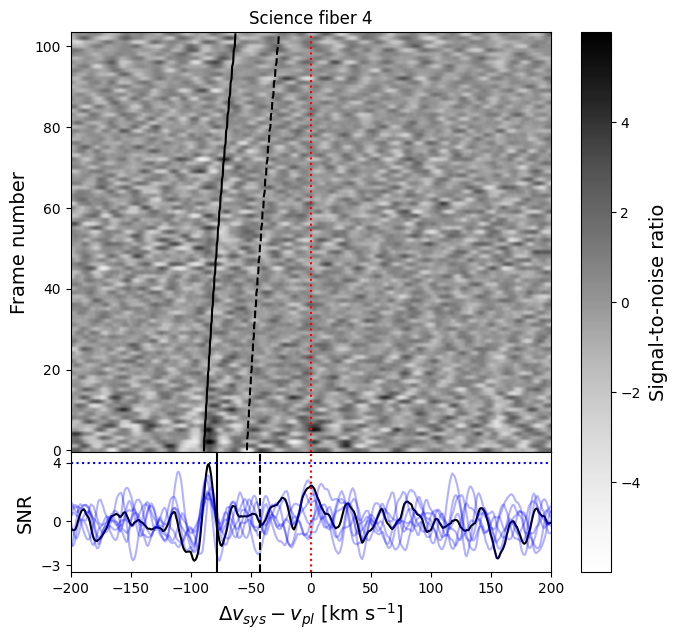}
    \caption{Top: Frame-by-frame signal-to-noise for the maximum likelihood model as a function of velocity offset in the planet reference frame. The stellar reference frame is indicated in solid black and the telluric reference frame in dashed black. The planet signal is expected along the dashed red line, but is not clear in either 2D time series. Bottom: Same as top, but summed over the time axis and re-normalized by the standard deviation in the $|\Delta v_{\rm sys} - v_{\rm pl}| > 25$ \kms\ region. Each individual order is shown in blue, and the combination of all orders is shown in black. A weak peak at SNR$\sim$2-3 is seen near the nominal planet reference frame in both time series, but does not dominate the space, and is too weak to be considered a detection. Contamination from stellar residuals can be seen in the first part of both time series}
    \label{fig:vtracks}
\end{figure*}

The retrieval analysis does not provide a clear confidence level for the planet detection, which we therefore assess using cross-correlation analysis. Figure \ref{fig:vtracks} shows the cross-correlation signal-to-noise for each frame with the retrieved maximum-likelihood model from the four component retrieval as a function of the velocity in the nominal planet rest frame, with the summed cross-correlation shown in the lower panels. A planet signal is expected to appear as a consistent peak in both fibers near $\Delta v_{\rm sys} - v_{\rm pl} = 0$ \kms. A weak ($\rm SNR\sim2-3$) peak consistent with the expected planetary ephemeris is seen in each fiber after summing over the time axis. Both fibers also show features near the stellar reference frame, particularly at the start of the observation sequence, indicating that some stellar residuals may be present in the early portion of the time series despite our detrending. This occurs when the velocity difference between planet and star is maximized, and therefore the risk of cross-contamination between stellar and planetary features is minimal. In both fibers, multiple orders are contributing to the overall signal, consistent with the presence of H$_2$O and CH$_4$ opacity features throughout the observed bandpass. Neither fiber individually produces a peak above the $\rm SNR\geq4$ threshold for a tentative HRCCS detection.

\begin{figure*}
    \centering

    \includegraphics[width=0.3\linewidth]{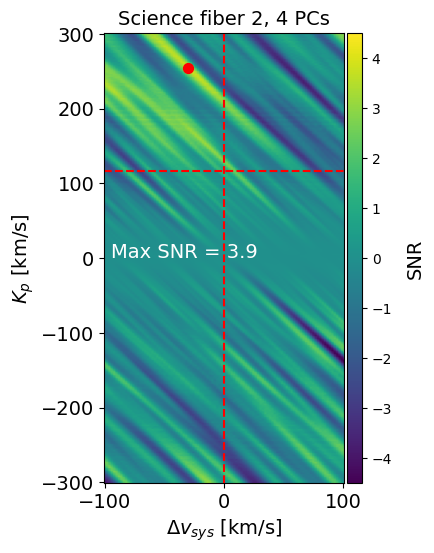}
    \includegraphics[width=0.3\linewidth]{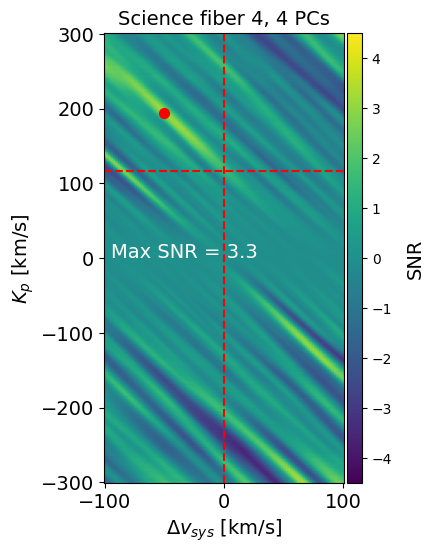}
    \includegraphics[width=0.3\linewidth]{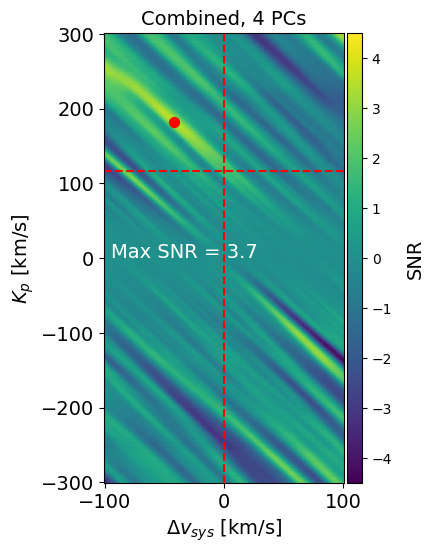}

    \includegraphics[width=0.3\linewidth]{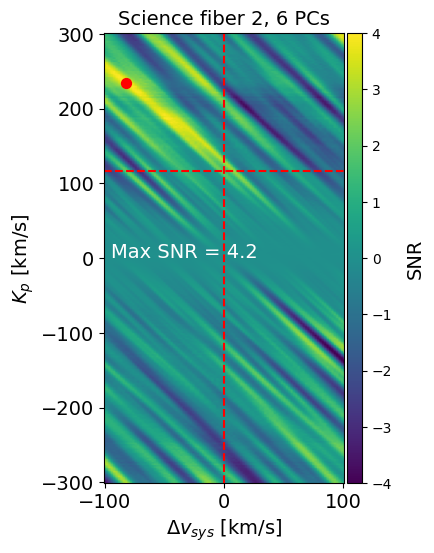}
    \includegraphics[width=0.3\linewidth]{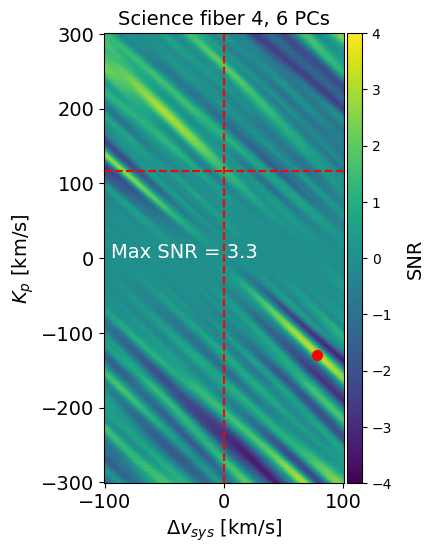}
    \includegraphics[width=0.3\linewidth]{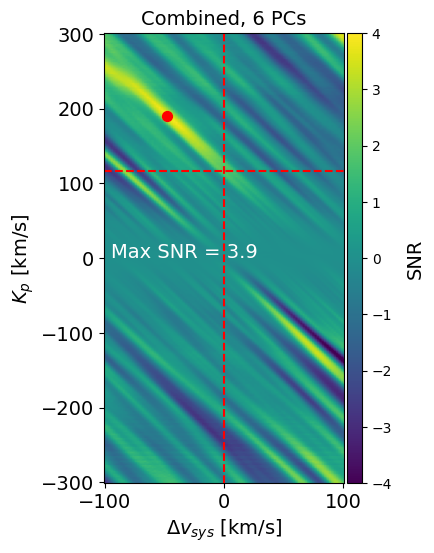}

    \includegraphics[width=0.3\linewidth]{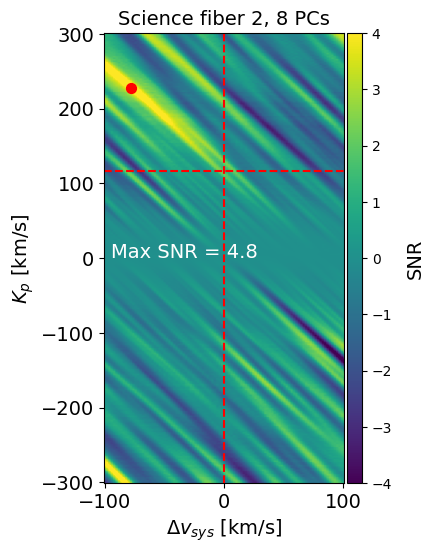}
    \includegraphics[width=0.3\linewidth]{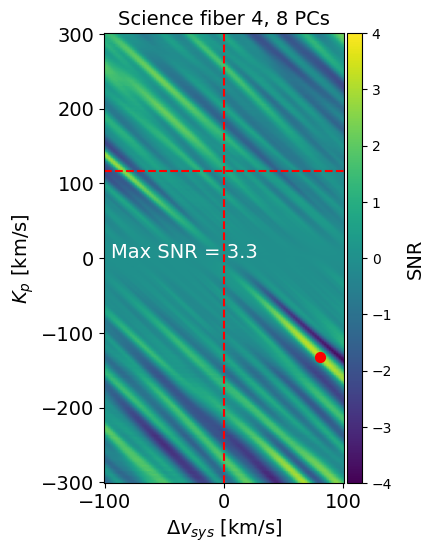}
    \includegraphics[width=0.3\linewidth]{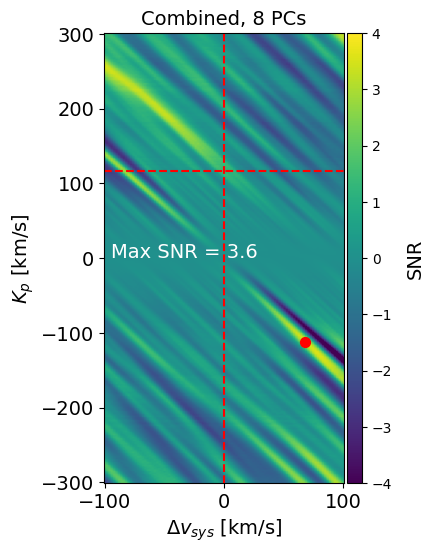}

    \caption{\kpvsys\ plots for science fiber 2 (left column), science fiber 4 (center) and both fibers (right) for 4 (top), 6 (middle), and 8 (bottom) omitted PCs, using the  maximum-likelihood planet model from the 4 component retrieval. The \kpvsys\ maps are computed using the log-likelihood for the grid of \kp\ and \dvsys, then converted to an SNR by median-subtracting each row of constant \kp\ and dividing by the standard deviation of the \kp$<0$ region. The expected $K_{\rm p}$ and \dvsys\ are indicated in dashed red, and the color bar is the same for all subplots. The red dot indicates the location of highest SNR. The possible planet detection is present at $\rm SNR\sim3-4$ in the science fiber 2 and combined data sets, while the science fiber 4 data shows a peak which is consistent in \kp\ and \dvsys\ but lower significance. In all cases, the potential peak is shifted to high $K_{\rm p}$ values.  Increasing the number of omitted components does not significantly change the \kpvsys\ map, consistent with the similar posteriors obtained from the three retrievals.}
    \label{fig:kpvsys_pcs}
\end{figure*}

Figure \ref{fig:kpvsys_pcs} shows the \kpvsys\ plots for the four component maximum-likelihood model for each fiber individually and for the sum of both fibers, for 4, 6, and 8 omitted PCs. The maps are computed using the \citet{brogi2019} log-likelihood function, then converted to a signal-to-noise ratio by first median-subtracting each row of constant \kp\ and then dividing by the standard deviation of the $K_{\rm p}<0$ region. The median subtraction is necessary to account for the impact of the model variance term in the \citet{brogi2019} log-likelihood function. This approach was used in our previous HRCCS analyses \citep{finnerty2024, finnerty2025a,finnerty2025b, finnerty2025c}. See \citet{finnerty2026a} for further details.

Combining the two time series results in a $\rm SNR\ \sim3-5$ peak regardless of the number of omitted PCs, with an offset compared with the expected location of any planetary features. The offset to higher $K_{\rm p}$ values is accompanied by more negative \dvsys\ values, such that the actual planet velocities are very similar to those expected from the assumed ephemeris. Based on negative injection tests at $-K_{\rm p}$, discussed in Section \ref{sec:disc}, this offset appears to be a result of the detrending process and/or orbital phase coverage of our observations. This potential planet peak is seen in both science fibers separately, consistent with the cross-correlation plots shown in Figure \ref{fig:vtracks}, but is significantly weaker in the science fiber 4 time series and degrades faster with increasing omitted PCs compared with the science fiber 2 time series. As the noise estimate in the \kpvsys\ map is determined from the map itself, summing the two maps results in a SNR intermediate to the two time series. 

Despite the higher noise level in the science fiber 4 time series, the consistency between the two time series in Figures \ref{fig:vtracks} and \ref{fig:kpvsys_pcs}  suggests that the putative planet peak is not simply a result of a one-off anomalous frame, which we would expect to produce a peak in the time series for only one fiber. The presence of the putative planet peak in the time series for each fiber at a generally similar strength suggests it is arising from real features in the observed spectra, rather than a purely noise-driven false positive. Other features in the \kpvsys\ space, particularly in the $K_{\rm p}<0$ region, are not as consistent between the two science fibers.

The potential planet feature is also consistent as the number of omitted principal components is changed. Previous HRCCS studies have demonstrated that false positives at SNRs similar to the possible planet feature can be introduced during the detrending process \citep[e.g.][]{cheverall2023}. In such a scenario, we would expect that the peak strength would vary strongly with the number of omitted PCs. Instead, Figure \ref{fig:kpvsys_pcs} shows that the strength of the potential planet feature decreases slowly with increasing the number of omitted PCs. This is consistent with expectations for a bona fide detection of \gjb\ with a small velocity baseline, resulting in rapid removal of the planet atmosphere with increasing the number of omitted PCs. While a secondary feature is stronger in the science fiber 4 time series for 6 and 8 omitted PCs, combining both science fibers continues to prefer the peak consistent with the expected planetary velocities in the 6 component case, albeit below the SNR$\geq4$ threshold generally assumed in HRCCS studies for a detection, and the offset peak is only weakly stronger in the 8 component case. This offset feature is consistent with the stellar reference frame, suggesting it may be related to the residual features apparent in Figure \ref{fig:vtracks}. 

\subsection{Chemical composition}\label{ssec:chem}

\begin{figure*}
    \centering
    
    \includegraphics[width=0.3\linewidth]{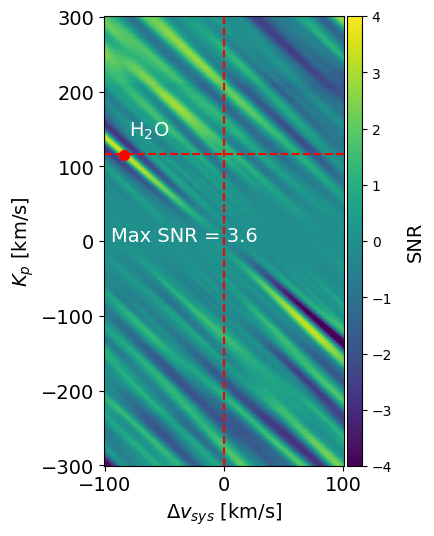}
    \includegraphics[width=0.3\linewidth]{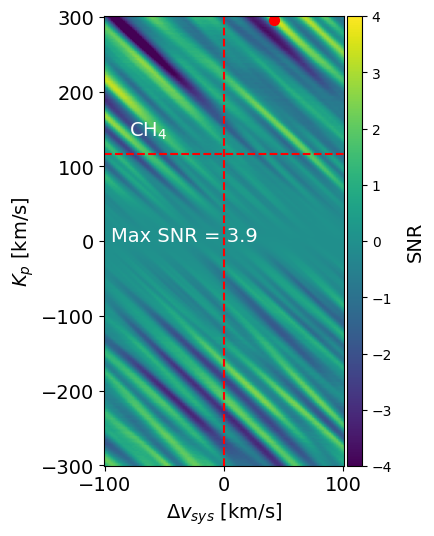}
    \includegraphics[width=0.3\linewidth]{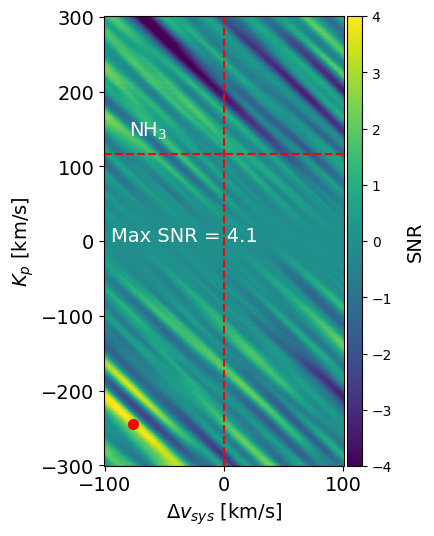}
    \includegraphics[width=0.3\linewidth]{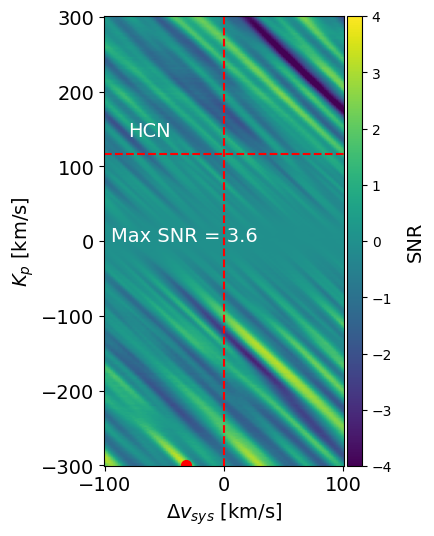}
    \includegraphics[width=0.3\linewidth]{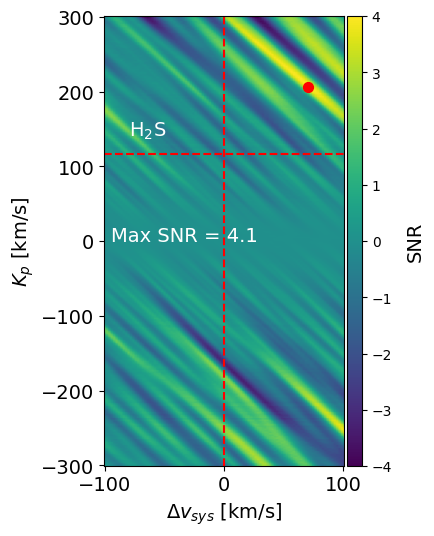}
    \includegraphics[width=0.3\linewidth]{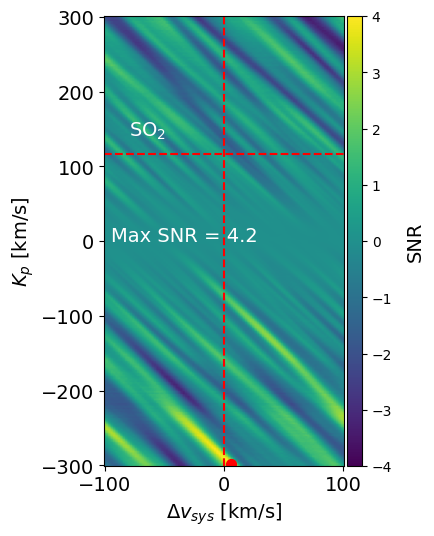}
    \caption{\kpvsys\ plots for each considered species. Templates were generated using the maximum-likelihood $P-T$ parameters, setting the VMR of the considered species to $10^{-1.5}$ and of that of all other species to $10^{-15}$. The \kpvsys\ map is generated and converted to an SNR map following the same procedure as Figure \ref{fig:kpvsys_pcs} The red dot indicates the location of the maximum SNR. The H$_2$O template produces a peak consistent with the potential planet peak at SNR$\sim3$, below the SNR$\geq4$ threshold for a tentative detection, while all other species produce little to no peak at the expected location in the \kpvsys\ space. This is consistent with the posterior preferring significant quantities of only H$_2$O, but with significant CH$_4$ opacity in the maximum-likelihood model contributing to the $\rm SNR = 3.9$ detection of the that model shown in Figure \ref{fig:kpvsys_pcs}.} The H$_2$O template also produces a significant feature consistent with the stellar reference frame, suggesting uncorrected stellar features remain in the detrended time series.
    \label{fig:kpvsys_mols}
\end{figure*}

Figure \ref{fig:kpvsys_mols} presents \kpvsys\ plots for each molecular species considered in the retrieval. For each species, we generated a template with the maximum-likelihood parameters, but set the volume-mixing ratio of the species in question to $10^{-1.5}$ and the VMR of the other species to $10^{-15}$. No individual species satisfies the $\rm SNR\geq4$ detection criteria individually, but H$_2$O produces an $\rm SNR\sim3$ peak consistent with the peak produced by the maximum-likelihood model. The significantly better detection of the maximum-likelihood model compared to the H$_2$O-only model is likely due to the relatively large CH$_4$ opacity in the maximum-likelihood model, but CH$_4$ is not interdependently detected, which is consistent with the retrieved posterior providing only an upper limit on the CH$_4$ abundance.  

The other species included in the retrieval all have substantially lower $L$ band opacities than H$_2$O or CH$_4$. Given the tentative nature of the overall planet detection, we would not expect to see evidence for these species unless they were the dominant constituents of \gjb's atmosphere in the region probed by our observations. Such a detection would indicate an atmosphere far from chemical equilibrium and/or would suggest a spurious detection of non-planetary features. A detection based primarily on CH$_4$ and H$_2$O features is consistent with expectations for the atmosphere of \gjb\ under chemical equilibrium. 

\subsection{Injection tests}

\begin{figure*}
    \centering
   \includegraphics[width=0.9\linewidth]{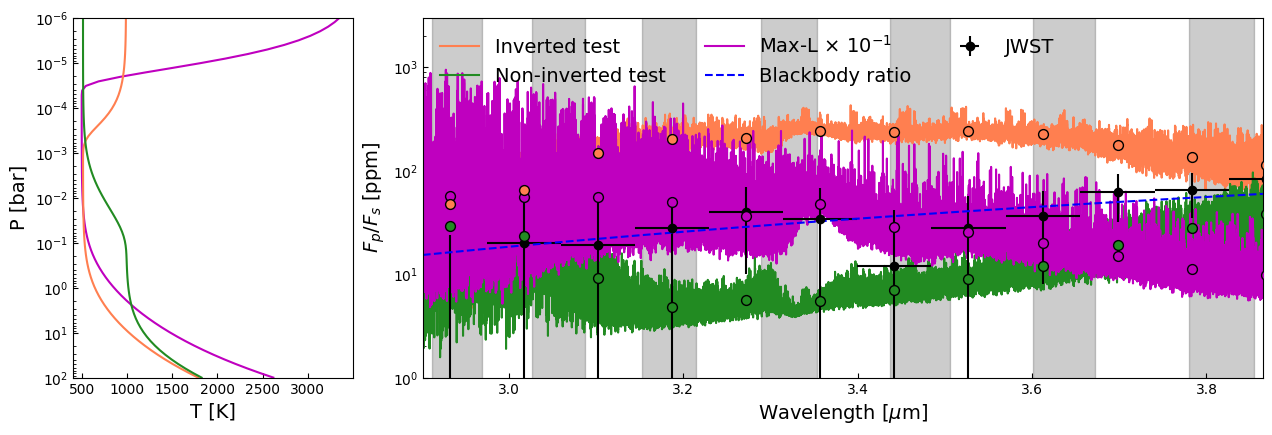}
    \caption{Left: $P-T$ profiles for the planet models used for injection tests. Right: planet spectra used for injection tests, compared with \textit{JWST} photometry form \citet{mukherjee2025}.}
    \label{fig:injmodels}
\end{figure*}

\begin{figure*}
    \centering
    \includegraphics[width=0.45\linewidth]{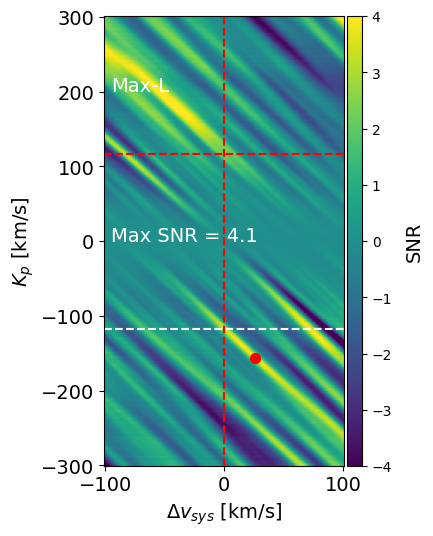}
    \includegraphics[width=0.45\linewidth]{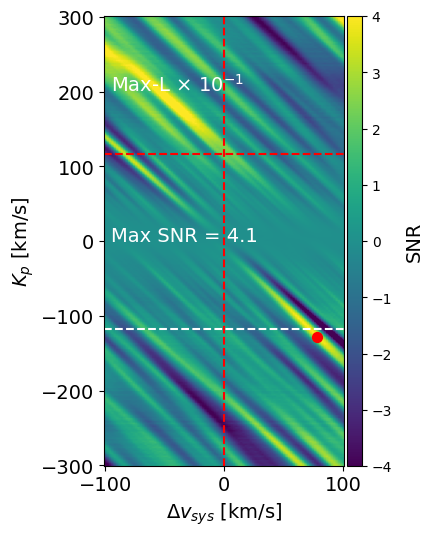}
    \includegraphics[width=0.45\linewidth]{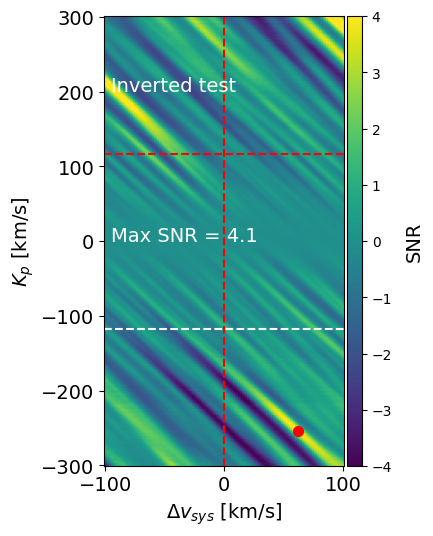}
    \includegraphics[width=0.45\linewidth]{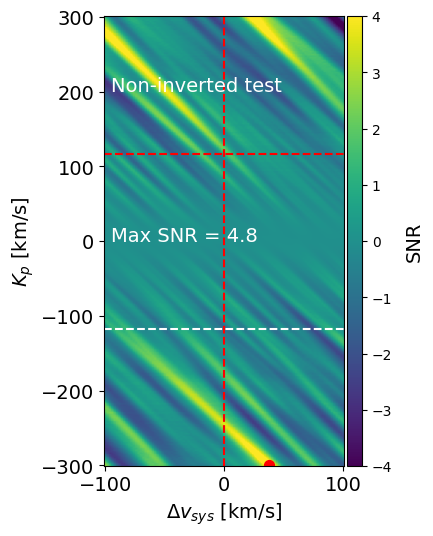}
    \caption{\kpvsys\ plots from injection at $-K_{\rm p}$ for the max-likelihood model (top left), max-likelihood multiplied by $10^{-1}$ (top right), an inverted test model (lower left) and non-inverted test model (lower right), omitting 4 principal components. The \kpvsys\ maps are computed similarly to Figure \ref{fig:kpvsys_pcs}, but the noise is estimated from the standard deviation of the region with both \kp$<0$ and \dvsys$<0$ to minimize contamination from the injected planet model. The red dot indicates the highest point in the \kp$<0$ region. The maximum-likelihood model is recovered at $\rm SNR=4.1$, slightly stronger than the tentative detection in Figure \ref{fig:kpvsys_pcs}, with a similar velocity offset. None of the other models are recovered, suggesting the molecular features in these models are too weak to be detectable in these data.
    For the maximum-likelihood $\times10^{-1}$ case, the tentative planet feature in the positive \kp\ region is consistent with the equivalent plot for the full-flux max-likelihood model, while the injected model is not recovered. This demonstrates that the SNR in the \kpvsys\ space is independent of the planet continuum level, but that injection-recovery tests are sensitive to the strength of the injected planet features.}
    \label{fig:neginj}
\end{figure*}

To assess the robustness of the detected signal, we performed a series of injection tests, where we inject a planet model onto the observed data at $-$\kp\ and then compute the \kpvsys\ diagram. We consider the maximum-likelihood model from the retrieval, the maximum-likelihood model scaled by $10^{-1}$ to better match the \textit{JWST} photometry \citep{mukherjee2025}, a model with a weaker thermal inversion than the retrieved maximum likelihood model, intended to mimic a more realistic inversion arising above a cloud deck in a high-metallicity atmosphere, and a non-inverted, high-metallicity model. The molecular abundances for the test models are based on equilibrium abundances from \texttt{easyCHEM} \citep{lei2024} for the assumed $P-T$ profiles with C/O = 0.6 and [Fe/H] = 2.0.  The parameters for these test models are included in Table \ref{tab:priors}, and the $P-T$ profiles and spectra are plotted in Figure \ref{fig:injmodels}, with the \citet{mukherjee2025} \textit{JWST} photometric points for comparison. The resulting \kpvsys\ maps are shown in Figure \ref{fig:neginj}. These tests give a sense of what planet signals could plausibly be recovered from these data.  

The maximum-likelihood model is recovered at $\rm SNR = 4.1$, while the other models are not recovered. The maximum-likelihood model is recovered with a substantial offset to higher magnitudes of \kp, with a higher value of \dvsys\ countering this to keep the planet velocities similar to the values of the expected ephemeris. As the planet was injected at $(0,-K_{\rm p})$, this offset must be a result of our data processing and/or the phase coverage of our observations. This offset is also seen in Figures \ref{fig:kpvsys_pcs} and \ref{fig:kpvsys_mols}, and is of roughly similar magnitude, suggesting that the apparent offsets in the \kpvsys\ maps for the maximum-likelihood models are consistent with a true planet signal at $(0,K_{\rm p})$. 

This offset is likely related to the relatively small velocity shift of our observations (26 \kms\ compared with a $10-14$ \kms\ instrumental resolution) leading to self-division of the planet model when the data-detrending procedure is replicated on the forward model. Self-division decreases as $K_{\rm p}$ increases, and does not seem to be a major factor for velocity shifts $\gtrsim 5\times$ the instrumental resolution. For smaller velocity shifts, self-division results in the variance of the planet model increasing with $K_{\rm p}$. The \citet{brogi2019} log-likelihood function depends directly on the model variance, leading to a systematic change in the log-likelihood with $K_{\rm p}$:

\begin{equation}\label{eq:logl}
    \log \mathcal{L} = -\frac{N}{2}\log \left[ \frac{1}{N}\sum(f^2 + g^2 - 2fg) \right] 
\end{equation}

Where $f$ is the data, $g$ is the planet model, and $N$ is the number of pixels, and the $g^2$ term is the variance of the forward model. If there is significant self-division of the planet model, $g^2$ rapidly goes to 0 as the self-division increases, leading to an increase in the log-likelihood for lower magnitudes of $K_{\rm p}$, but a higher variance in the log-likelihood values at higher $K_{\rm p}$ as $g$ is less impacted by division and the potential magnitude of $fg$ becomes larger. 

This effect can be clearly seen in all of our \kpvsys\ plots. While the cross-correlation coefficient does not directly depend on the model variance, the overall amplitude of the planet model still changes with \kp, reaching 0 for \kp$=0$, and this effect on the forawrd model will still be present. The resulting increase in the maximum obtainable CCF value with $K_{\rm p}$ will therefore lead to a bias towards higher $K_{\rm p}$ values when self-subtraction of the planet model is significant, as is the case in these data.

The injected maximum-likelihood model is recovered at a similar strength to the potential planet feature in Figure \ref{fig:kpvsys_pcs}, indicating that the detection of such a spectrum is plausible given the quality of our data. In contrast, other three models are not recovered, which suggests that the spectral features of these models are too weak relative to the overall continuum level to be detected given the quality of these data. Critically, it is the strength of the planet features relative to the stellar continuum level which is of importance cross-correlation analysis, rather than the planet continuum level. We discuss this in detail in Section \ref{sec:disc}. The implication is that the maximum-likelihood model is matching the strength of line features in the observations, but the lack of continuum sensitivity, coupled with the degeneracy between the $P-T$ profile contrast and molecular abundances, limits our ability to understand what is physically setting the strength of these spectral features. 

\subsection{Retrieval at $-\rm K_p$}

Finally, we performed a retrieval omitting 4 principal components, but with the reference value of \kp\ set to $-$\kp, i.e. $-117$ \kms\ rather than 117 \kms. While ideally, a retrieval in the absence of statistically significant planetary features would return a posterior consistent with the assumed priors, in practice the sampler will attempt to fit any features in the data, which may result in the retrieval converging on uncorrected stellar/telluric residuals or noise features. Performing a retrieval in the $-$\kp\ region, where we do not expect any planet signal, allows us to assess the behavior of the sampler in the absence of a planet.

The resulting posterior is generally similar to the retrieved posterior from the fiducial retrieval omitting four principal components, preferring a spectrum dominated by strong H$_2$O emission features. While this similarity is initially concerning, the retrieved \kp\ is approximately $-65$ \kms\, and retrieved \dvsys\ is $36$ \kms. These values result in a planet velocity track which overlaps the stellar reference frame near the start of the time series. Figure \ref{fig:vtracks} shows that the maximum-likelihood model from the fiducial retrieval, which is also dominated by H$_2$O emission, produces a cross-correlation peak near the stellar reference frame at the start of the observation sequence, and the corresponding feature in the \kpvsys\ diagrams can be seen in Figure \ref{fig:kpvsys_pcs}. This suggests that in the absence of a planet signal, the retrieval is converging on uncorrected H$_2$O residuals in the stellar reference frame. While the fiducial retrieval prefers a similar spectrum to the retrieval with $-$\kp, the preferred reference frame which is more consistent with \gjb\ than the host star, indicating the fit is not being driven by features in the host star. However, this test raises the troubling prospect of non-planetary features producing spectrum which appears similar to the fiducial retrieved spectrum, which is an important caveat to this analysis. 

\section{Discussion}\label{sec:disc}

\begin{figure*}
    \centering
    \includegraphics[width=0.95\linewidth]{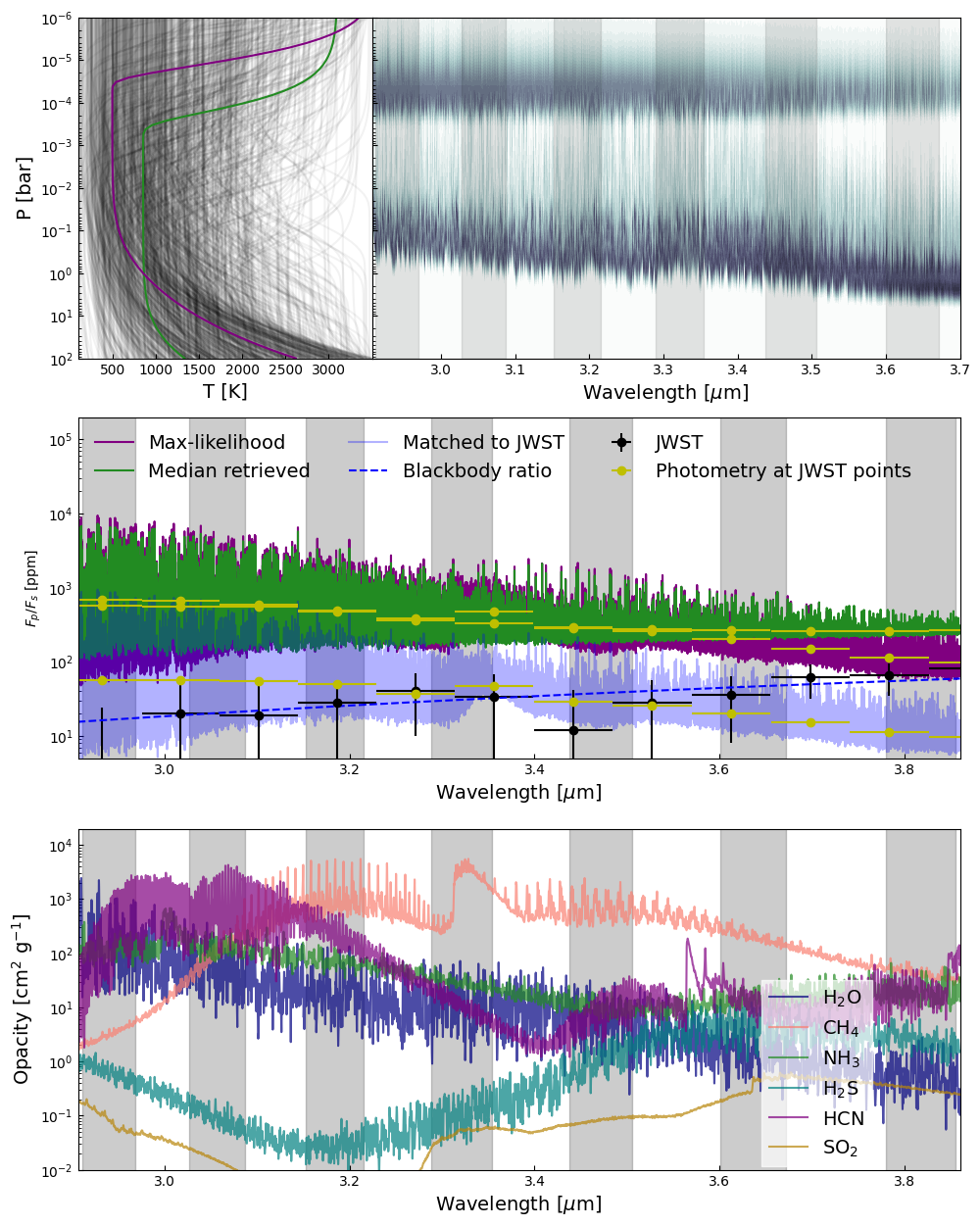}  
    \caption{From top left: Retrieved $P-T$ profiles, maximum-likelihood emission contribution function, model $F_p/F_s$, and opacities. Observed orders are shaded in gray. The $F_p/F_s$ includes the \textit{JWST} low-resolution emission points from \citet{mukherjee2025}, and also the expected $F_p/F_s$ assuming the planet and star are 670 K and 3600 K blackbodies, respectively. For each model, we show the photometry at the \textit{JWST} points in yellow.}
    \label{fig:specplot}
\end{figure*}

\subsection{Insensitivity to continuum}

A key factor in the interpretation of our results is the insensitivity of HRCCS techniques to the continuum level of the planet. During data processing, the spectral time series for each order is divided by its own median in order to remove static stellar and telluric features. In the case of a rapidly moving planet, the planet signal is effectively absent from the resulting time series, which is simply $F_p/F_s$ after the median division. However, if the observed planet velocity shift is only $2-3\times$ the instrument resolution,  there will be some planet contribution to the median spectrum, and the denominator will contain a contribution from the planet spectrum as well as the stellar spectrum. The impact of this is minimized by processing the forward model in the same way in order to replicate any such self-division effects, but an overall reduction in the strength of the planet signal will still degrade detectability. The comparison between the model and data is typically done using the cross-correlation coefficient or a log-likelihood mapping such as the one described in \citet{brogi2019}, which can be re-written in terms of the cross-correlation coefficient. For an observed spectrum $f$, forward model $g$, and their associated variances $\sigma^2_{f}$ and $\sigma^2_g$, the cross-correlation is:

\begin{equation}\label{eq:cc}
    C = \frac{\sum fg}{N\sqrt{\sigma^2_f\sigma^2_g}}
\end{equation}

Where $N$ is the number of points in the spectra and the sum is carried over a spectral order. The forward model $g$ can be Doppler-shifted to different velocities and $C$ recalculated in order to obtain $C(v)$, and we expect that a planet detection will appear as a peak at $C(v=v_{pl})$, where $v_{pl}$ is the expected radial velocity of the planet based on a prior ephemeris.

A major issue in early cross-correlation analyses is that Equation \ref{eq:cc} is scale-invariant. Multiplying $g$ by a constant positive value results in no change to $C$, as $\sqrt{\sigma_g^2}$ is changed by the same factor. This results in the top panels of Figure \ref{fig:vtracks} being unchanged if the planet model is scaled up or down by a factor of 10, as can be seen in the top two panels of Figure \ref{fig:neginj}. This mathematical shortcoming is addressed by log-likelihood mappings such as \citet{brogi2019} (Equation \ref{eq:logl}), which have enabled atmospheric retrievals. Even with these mappings, observations over a short bandpass have struggled to retrieve accurate absolute temperatures as a result of degeneracies between scaling factors and $P-T$ profile parameters \citep[e.g.][]{finnerty2023, finnerty2025b}. Wider bandpasses become more sensitive to slow changes in $F_p/F_s$ which cannot be well-described by a multiplicative scaling factor, resulting in better constraints on absolute temperature \citep[e.g.]{line2021, finnerty2025c}. In the case of \gjb\ in the $L$ band, $F_p/F_s$ changes very slowly, suggesting our observations will have very poor sensitivity to the absolute flux level. Moreover, since the noise in the \kpvsys\ map is generally estimated from the map itself, the sensitivity of the log-likelihood function to overall scaling of the planet is effectively divided out, as can be seen in comparing the top panels of Figure \ref{fig:neginj}.

Figure \ref{fig:specplot} shows that the maximum-likelihood model has $F_p/F_s\approx10\times$ the expected values from the \citet{mukherjee2025} analysis of \textit{JWST} secondary eclipse observations, which is consistent with treating the star and planet as blackbodies. This would seem to suggest that the maximum-likelihood model is not representative of a planet detection. However, as Figures \ref{fig:specplot} and \ref{fig:neginj} show, multiplying the maximum-likelihood model by $10^{-1}$ gives a flux level roughly compatible with \citet{mukherjee2025} in the $L$ band and a detection strength in the \kpvsys\ space identical to the original maximum-likelihood model.

While this result appears to contradict the injection tests shown in Figure \ref{fig:neginj}, the insensitivity of HRCCS analysis to mismatches in scale between the data and the template does not apply in the case of injection tests. Rather than scaling one of the data or template by a multiplicative factor, these tests add a synthetic planet model $m$ to the observed spectrum $f$:

\begin{equation}\label{eq:ccinj}
    C = \frac{\sum (f+m)g}{N\sqrt{\sigma^2_{f+m}\sigma^2_g}} = \frac{1}{N\sqrt{\sigma^2_{f+m}\sigma^2_g}}\left[\left( \sum fg \right) + \left(\sum mg\right)\right]
\end{equation}

Applying a multiplicative factor to $m$ directly changes the relative contribution of the two terms in the sum, and the result is therefore sensitive to the strength at which the model is injected. 

The result of this can be seen in the upper right panel of Figure \ref{fig:neginj}, which shows the \kpvsys\ map obtained from injecting the maximum-likelihood model at $-$\kp\ at $0.1\times$ nominal strength and then cross-correlating with the maximum-likelihood model $\times0.1$. The \kp$>0$ region of this map appears to be identical to the map obtained from using the maximum-likelihood model at nominal strength in the top left panel of Figure \ref{fig:neginj}. In contrast, there are no significant features in the $-$\kp\ region near where the planet was injected at $0.1\times$ nominal strength, but injecting at nominal strength results in a feature of similar strength to the potential planet feature in the \kp$>0$ region.

These results suggest that the maximum-likelihood model is correctly matching the strength of features in the data relative to the stellar continuum, which are apparently substantially stronger than the test models considered. However, this still does not provide direct information on the continuum level of the planet, but rather constrains the strength of planetary lines relative to the total star+planet continuum, which is dominated by the stellar contribution. Planet lines which are strong relative to the continuum level of the planet can produce a comparable signal to weaker lines (relative to the planet continuum) if the planet continuum in the latter case is larger, such that the strength of the planet lines relative to the host star is consistent. Such a scenario impacting our retrievals is necessary to reconcile these results with the \textit{JWST} spectrum from \citet{mukherjee2025}.  This could be due to differences in the $P-T$ profile or very high molecular abundances, which could increase line strengths while maintaining a continuum in agreement with \citet{mukherjee2025}. 

This overall insensitivity to scale also provides an explanation for the apparent $\rm SNR\sim4$ feature near the planet reference frame with the non-inverted test model in the bottom right corner of Figure \ref{fig:neginj}. While this result appears to be in conflict with the maximum-likelihood model preferring a significant inversion, examination of the actual $\Delta \log L$ values shows the non-inverted test model produces a much smaller peak than the maximum-likelihood model. However, the reduction in the template variance also generally reduces the scale of the $\log L $ variations, and therefore the estimated noise level. The net result is that the small $\log L$ peak near the planet reference frame appears to have a comparable signal-to-noise to the much higher likelihood inverted models, because the change in the overall scale is effectively divided out. The failure to recover the non-inverted model when injected at $-$\kp\ confirms that detection of this model is unlikely given the noise level of the observations.

\subsection{Thermal structure and clouds}

Previous observations of \gjb\ have favored a non-inverted, high-metallicity atmosphere \citep{morley2017, lothringer2018GJ, grasser2024, mukherjee2025}. Support for clouds and hazes has been more mixed, with \citet{morley2017} not seeing significant evidence for either, \citet{lothringer2018GJ} requiring hazes for metallicities $\sim100\times$ solar but not for higher metallicities, and \citet{grasser2024} favoring a cloud deck with $\rm P < 10\rm\ mbar$ and metallicity $>300\times$ solar. Most recently, \citet{mukherjee2025} favored models with clouds and metallicity $>300\times$ solar, but also found that cloud-free atmospheres with photochemistry and metallicity $\sim80\times$ solar are compatible with the \textit{JWST} eclipse observations. Both the high-resolution observations presented by \citet{grasser2024} and the \textit{JWST} observations presented by \citet{mukherjee2025} independently support the cloudy, high-metallicity interpretation, and neither analysis is reliant upon the controversial \textit{Spitzer} photometry.

In contrast, our analysis prefers an inverted $P-T$ profile, and the retrieval significantly overestimates the planet flux relative to the values reported by \citet{mukherjee2025}, which are consistent with a blackbody at the nominal equilibrium temperature of \gjb. Non-inverted profiles are rejected at 95.7\%, 99.0\%, and 57.8\% confidence for the 4, 6, and 8 component retrievals respectively. The decrease in confidence is consistent with the overall broadening of the posterior, likely due to removal of planet features as the number of omitted PCs is increased. 

While the retrievals prefer an inversion, the constraints on the $P-T$ profile and corresponding planet flux from these high-resolution data are relatively loose, and there is only a weak constraint on the pressure level or strength of the inversion (see Figure \ref{fig:specplot}). While the maximum-likelihood model prefers a very strong inversion at $\sim10\ \mu\rm bar$, the median model prefers a somewhat deeper and weaker inversion, and the posterior draws shown in Figure \ref{fig:specplot} include profiles with inversions as deep as 1 bar. The analysis of \textit{JWST} observations presented in \citet{mukherjee2025} permits inversions above $\sim1\ \rm mbar$ (see their Figure 5), but with a much smaller $P-T$ contrast than the 2500 K of the maximum-likelihood model. Planetary lines with a strength similar to the retrieved maximum-likelihood model but a $P-T$ profile compatible with \citet{mukherjee2025} would require higher molecular abundances and a lower-altitude inversion than the maximum-likelihood values. 

In the case of GJ~1214~b, repeated theoretical studies have found that soot hazes can drive thermal inversions and significantly increase the planet flux in the $2-4\rm\ \mu m$ band compared with a blackbody \citep{morley2015, malsky2025, steinrueck2025}. GJ~1214~b orbits a similar host star to \gjb\ and receives similar irradiation, but has a significantly smaller mass ($\sim6\rm\ M_\odot$). The \citet{mukherjee2025} analysis preferred very low albedos for \gjb, potentially compatible with such a haze scenario. In \citet{malsky2025}, the soot haze models of GJ~1214~b show a $\sim300$ K thermal inversion, almost an order of magnitude smaller than the thermal contrast of our maximum-likelihood $P-T$ profile, which is plotted in Figure \ref{fig:specplot}. This suggests that our retrieval is substantially over-estimating the strength of the inversion, if indeed the retrieval is fitting the planetary atmosphere and an inversion is present. A weaker inversion combined with higher abundances could produce a spectrum with molecular features of similar strength and a continuum more in line with the results of \citet{mukherjee2025}. However, our retrieval setup assumes a H$_2$ dominated atmosphere with scattering opacity from $\rm H_2 -H_2$ and $\rm H_2-He$ interactions, an assumption which may begin to break down at high metallicities. If this is the case, the retrieved high-contrast $P-T$ profile and relatively low molecular abundances could provide a reasonable fit to the data while avoiding issues with the radiative transfer calculation at very high metallicities. 

While we included a gray cloud deck in our analysis, the posteriors for the cloud parameters are nearly flat. The achromatic gray cloud opacity is effectively impossible to disentangle from the scaling parameter and/or changes to the $P-T$ profile. Constraining clouds in emission with HRCCS will require broad wavelength coverage and more realistic cloud models to infer the impact of clouds without relying solely on the planet continuum. 

\subsection{Atmospheric composition}

The poor sensitivity of HRCCS analysis to the planetary continuum level creates a degeneracy between thermal contrast in the $P-T$ profile and molecular abundances, as both larger thermal contrast and greater abundances increase the strength of molecular features. The retrieved median parameters has a very large thermal contrast ($\sim2000$ K), but relatively low abundances compared to the values expected for a bulk $[\rm Fe/H]\sim2$. Atmospheres with larger abundances but lower $P-T$ contrast are in better agreement with previous observations (see Figure \ref{fig:injmodels}), which motivated our parameter choices for the inverted test model (listed in Table \ref{tab:priors}). However, this model is not recovered in the injection-recovery tests shown in Figure \ref{fig:neginj}, indicating that the planet lines in this model are too weak to be plausibly detected in these data. Stronger lines would require either higher abundances or greater $P-T$ contrast, the latter of which is the case for the maximum-likelihood model. 

% While this model is recovered in the injection-recovery tests shown in Figure \ref{fig:neginj}, it produces only a marginal cross-correlation peak in the data directly, suggesting that the model features differ significantly from the features which are being fit by the maximum-likelihood model. 

Based on Figure \ref{fig:kpvsys_mols}, the tentative detection is a result of a combination of H$_2$O and CH$_4$ features. Atmospheric opacity dominated by CH$_4$ and H$_2$O features is consistent with chemical equilibrium predictions. While previous analyses including \textit{Spitzer} photometry have preferred a high CO/CH$_4$ ratio based on the 3.6 $\mu$m flux, suggesting disequilibrium chemistry, this is in tension with results from \citet{mukherjee2025}, who found the high \textit{Spitzer} 3.6 $\mu$m flux could be a result of instrumental systematics. 

% \subsection{Comparison to previous results}
% Have a subsection on what it means if its a false positive?

\section{Summary and Conclusions}\label{sec:conc}

We present Keck II/KPIC emission observations of the warm Neptune \gjb\ in the $L$ band, covering $\sim2.91-3.85\rm\ \mu m$. While the observed velocity shift was only $\sim2\times$ the instrument resolution, cross-correlation analysis yields a tentative detection consistent with the expected reference frame of the planetary atmosphere. Our retrieval analysis returns a constrained posterior after omitting four principal components, and compatible but broader posteriors when omitting six or eight components. The preferred model from the retrieval has a strong thermal inversion and opacity from H$_2$O and CH$_4$, potentially consistent with an atmosphere roughly in chemical equilibrium with soot hazes producing a thermal inversion.

However, the maximum-likelihood model overestimates the planet flux by a factor of $\sim10$ compared with recent results from \textit{JWST}. The relatively flat continuum level in the $L$ band presents a particular challenge for HRCCS retrievals, which are only indirectly sensitive to the continuum level. This is further compounded by the small velocity shift of our observations, which leads to shifts of the potential planet peak in the \kpvsys\ space in injection tests. Additional observations with wider orbital phase coverage are required to validate this potential detection. 

This work demonstrates the potential of high-stability, high-resolution $L$ band spectroscopy for characterizing smaller and cooler exoplanets via emission spectroscopy. KPIC was decommissioned at the end of the 2025B observing semester, precluding additional observations with this instrument, and the increased thermal background associated with slit-fed instruments will likely preclude similarly sensitive observations in this bandpass with any existing instrument. Moving forward, warm Neptune characterization is a compelling science case to drive the development of new fiber-fed high-resolution spectrometers operating in the thermal infrared.   

\section*{acknowledgments}
We thank the anonymous referee whose thoughtful comments improved the quality of this paper. 
L. F. was a member of UAW local 4811 while performing this work, and is currently a member of UM-PRO. L.F. acknowledges the support of the W.M. Keck Foundation, which also supports development of the KPIC facility data reduction pipeline. The contributed Hoffman2 computing node used for this work was supported by the Heising-Simons Foundation grant \#2020-1821. Funding for KPIC has been provided by the California Institute of Technology, the Jet Propulsion Laboratory, the Heising-Simons Foundation (grants \#2015-129, \#2017-318, \#2019-1312, \#2023-4597, \#2023-4598), the Simons Foundation (through the Caltech Center for Comparative Planetary Evolution), and the NSF under grant AST-1611623. D.E. acknowledges support from the NASA Future Investigators in NASA Earth and Space Science and Technology (FINESST) fellowship under award \#80NSSC19K1423, as well as support from the Keck Visiting Scholars Program (KVSP) to install the Phase II upgrades for KPIC. J.X. acknowledges support from the NASA Future Investigators in NASA Earth and Space Science and Technology (FINESST) award \#80NSSC23K1434.

This work used computational and storage services associated with the Hoffman2 Shared Cluster provided by UCLA Institute for Digital Research and Education’s Research Technology Group. L.F. thanks Briley Lewis for her helpful guide to using Hoffman2, and Paul Molli\`ere for his assistance in adding additional opacities to petitRADTRANS. 

The data presented herein were obtained at the W. M. Keck Observatory, which is operated as a scientific partnership among the California Institute of Technology, the University of California and the National Aeronautics and Space Administration. The Observatory was made possible by the generous financial support of the W. M. Keck Foundation. W. M. Keck Observatory access was supported by Northwestern University and the Center for Interdisciplinary Exploration and Research in Astrophysics (CIERA). The authors wish to recognize and acknowledge the very significant cultural role and reverence that the summit of Mauna Kea has always had within the indigenous Hawaiian community.  We are most fortunate to have the opportunity to conduct observations from this mountain. 

This research has made use of the NASA Exoplanet Archive \citep{10.26133/nea12} and the Exoplanet Follow-up Observing Program \citep{10.26134/ExoFOP5}, which are operated by the California Institute of Technology, under contract with the National Aeronautics and Space Administration under the Exoplanet Exploration Program. The research shown here acknowledges use of the Hypatia Catalog Database, an online compilation of stellar abundance data as described in Hinkel et al. (2014, AJ, 148, 54), which was supported by NASA's Nexus for Exoplanet System Science (NExSS) research coordination network and the Vanderbilt Initiative in Data-Intensive Astrophysics (VIDA).

\software{astropy \citep{astropy:2013, astropy:2018},  
          \texttt{corner} \citep{corner},
          \petit\ \citep{prt:2019, prt:2020}}

\vspace{5mm}
\facilities{Keck:II(NIRSPEC/KPIC)}




\appendix
\section{Corner Plots}\label{app:corner}

\begin{figure}
    \centering
    \includegraphics[width=1.0\linewidth]{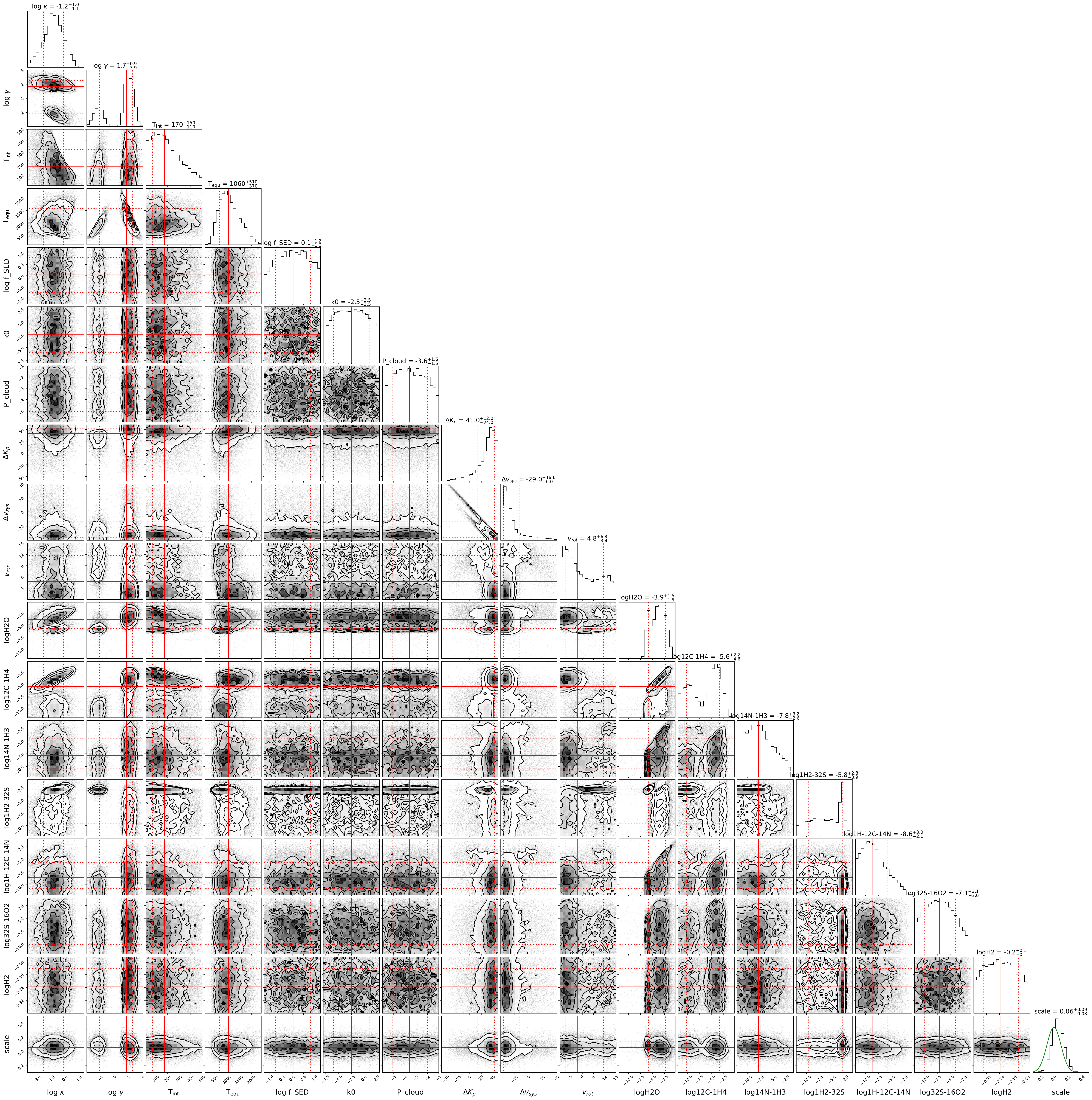}
    \caption{Full corner plot for the 2-component retrieval.}
    \label{fig:corner2}
\end{figure}

\begin{figure}
    \centering
    \includegraphics[width=1.0\linewidth]{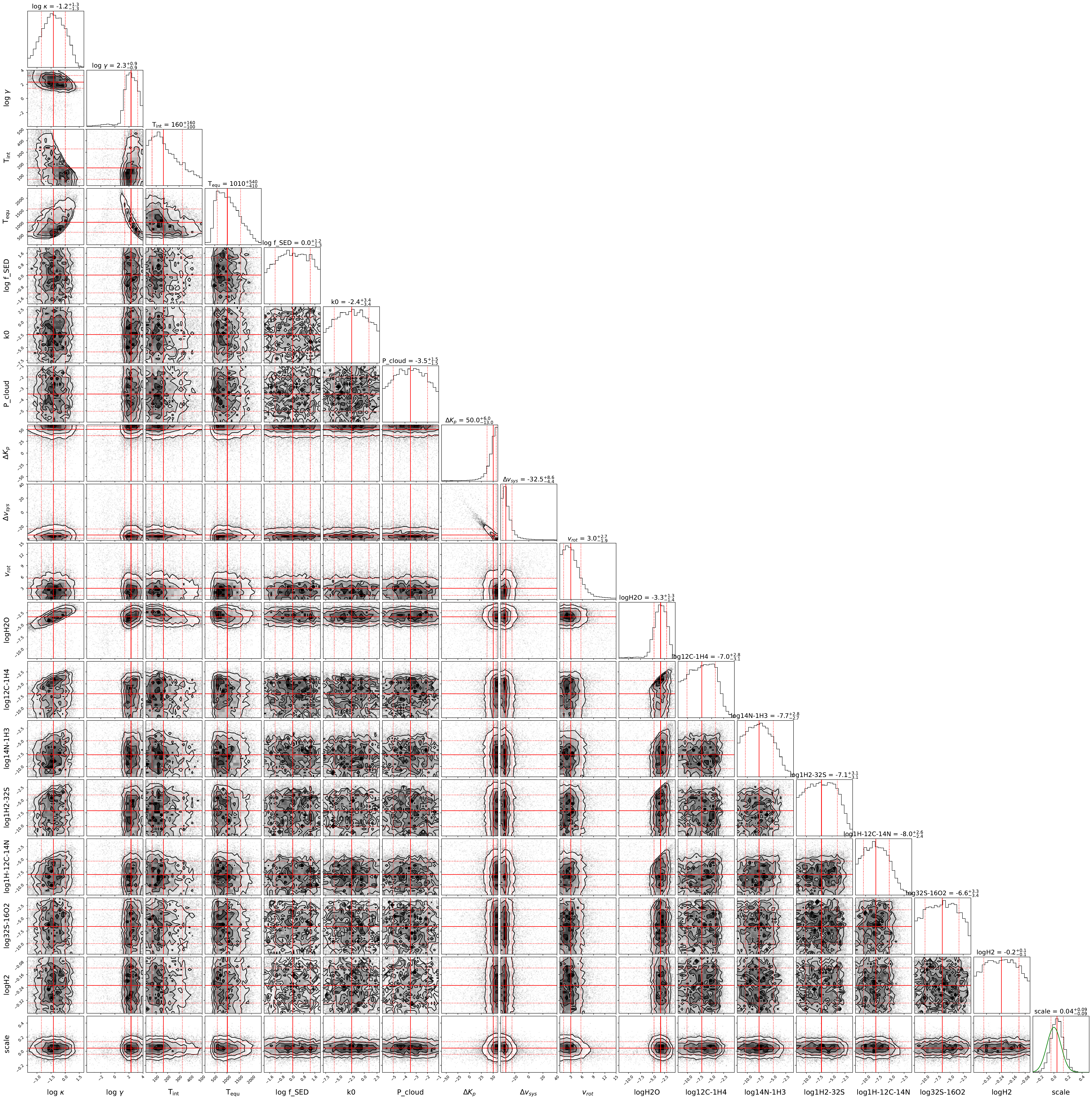}
    \caption{Full corner plot for the 4-component retrieval.}
    \label{fig:corner4}
\end{figure}

\begin{figure}
    \centering
    \includegraphics[width=1.0\linewidth]{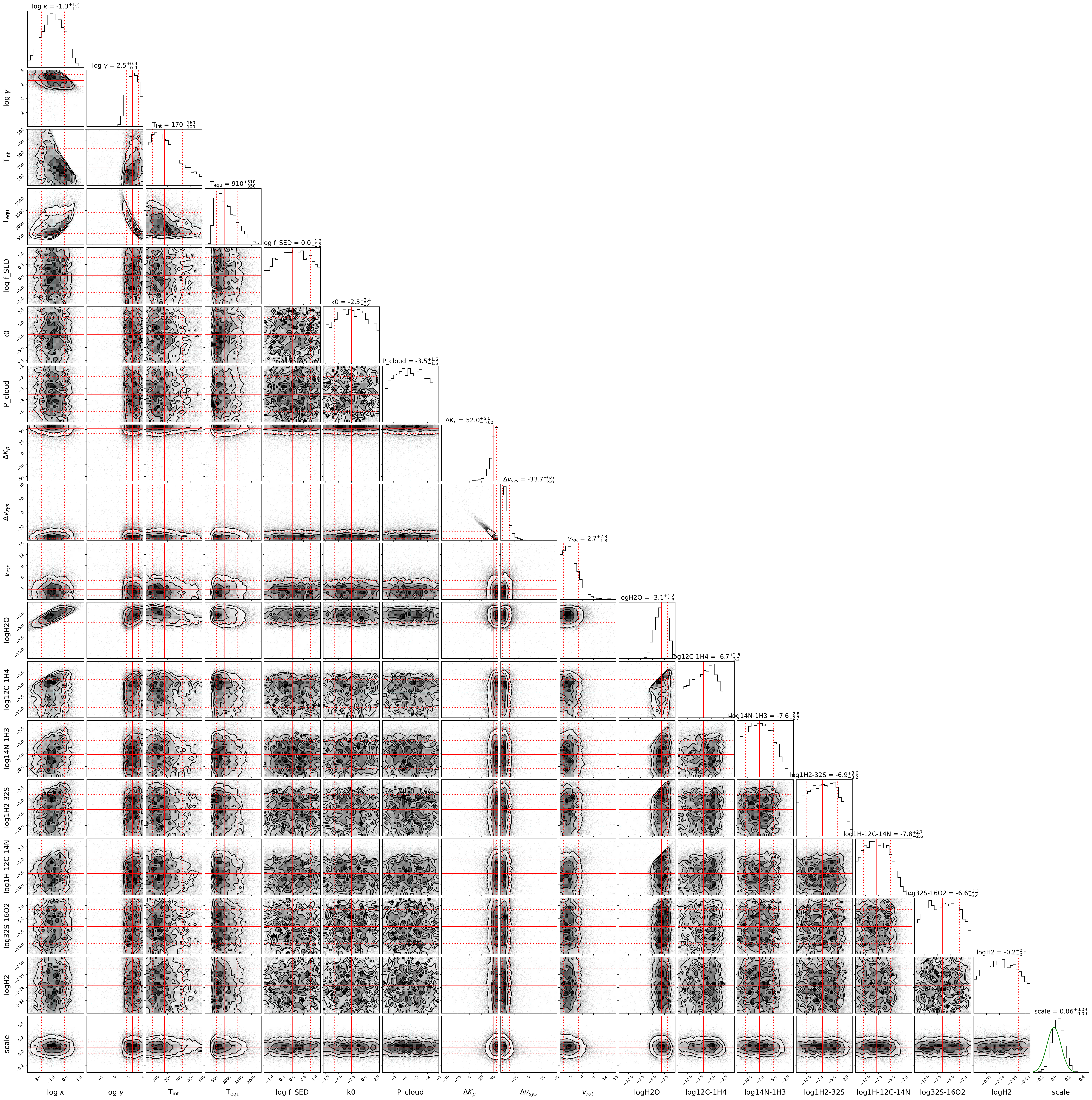}
    \caption{Full corner plot for the 6-component retrieval.}
    \label{fig:corner6}
\end{figure}

Figure \ref{fig:corner2}, \ref{fig:corner4}, and \ref{fig:corner6} present the full corner plots for the retrievals omitting 2, 4, and 6 principal components, respectively. 

\begin{figure}
    \centering
    \includegraphics[width=1.0\linewidth]{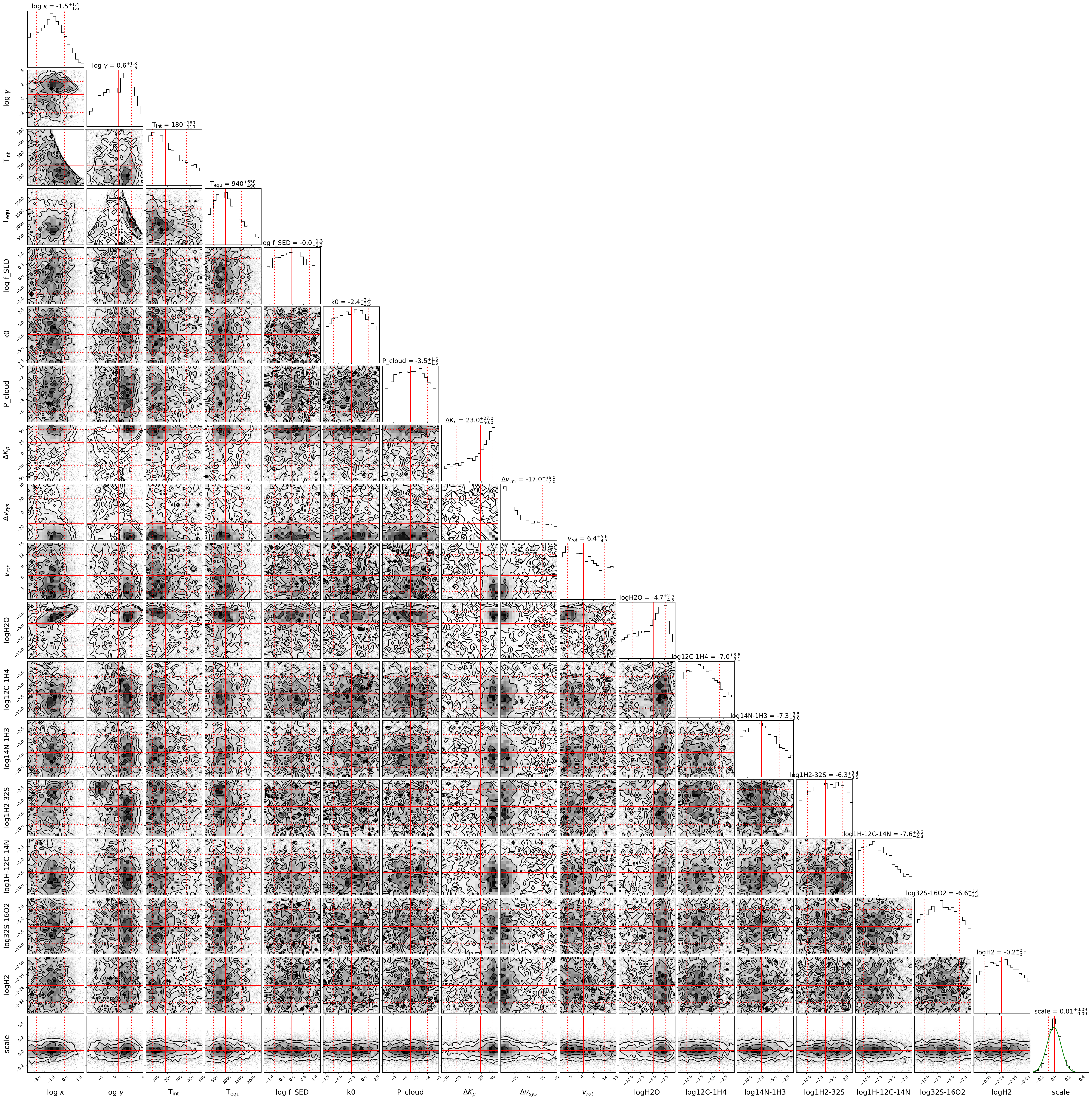}
    \caption{Full corner plot for the 8-component retrieval.}
    \label{fig:corner8}
\end{figure}

\clearpage
\bibliography{exoplanetbib}{}
\bibliographystyle{aasjournal}



\end{CJK*}
\end{document}